\documentclass[conference]{IEEEtran}

\usepackage{mathptmx}
\usepackage{biblatex}
\usepackage{csquotes}
\usepackage{graphicx}
\usepackage{listings}
\usepackage{orcidlink}
\usepackage{subcaption}
\usepackage{xcolor}
\usepackage{url}
\usepackage{hyperref}
\usepackage[capitalize,noabbrev,nameinlink]{cleveref} 

\newcommand{\op}[1]{\hat{#1}}

\begin{document}

\title{The Quantum Hamiltonian Analysis Toolkit:\\Lowering the Barrier to Quantum\\Computing with Hamiltonians}

\author{
    \IEEEauthorblockN{
        Brendan K.\ Krueger \orcidlink{0000-0002-8275-9277}\IEEEauthorrefmark{1},
        Stephan Eidenbenz \orcidlink{0000-0002-2628-1854}\IEEEauthorrefmark{2},
        Shamminuj Aktar \orcidlink{0000-0001-5587-7406}\IEEEauthorrefmark{2}
        Rishabh Bhardwaj \orcidlink{0000-0002-4979-0969}\IEEEauthorrefmark{2},\\
        John K.\ Golden \orcidlink{0000-0001-9369-0925}\IEEEauthorrefmark{2},
        George Grattan \orcidlink{0009-0004-2173-6970}\IEEEauthorrefmark{2},
        Abhijith Jayakumar \orcidlink{0000-0001-5297-8033}\IEEEauthorrefmark{3},
        Anna Matsekh \orcidlink{0000-0001-6274-904X}\IEEEauthorrefmark{4},\\
        Scott Pakin \orcidlink{0000-0002-5220-1985}\IEEEauthorrefmark{1},
        Nandakishore Santhi \orcidlink{0000-0002-4755-7821}\IEEEauthorrefmark{2},
        and
        Reuben Tate \orcidlink{0000-0002-9170-8906}\IEEEauthorrefmark{2},
    }
    \IEEEauthorblockA{\IEEEauthorrefmark{1}
        CAI-1: Applied Computer Science, Los Alamos National Laboratory, Los Alamos, NM, USA
    }
    \IEEEauthorblockA{\IEEEauthorrefmark{2}
        CAI-3: Information Sciences, Los Alamos National Laboratory, Los Alamos, NM, USA
    }
    \IEEEauthorblockA{\IEEEauthorrefmark{3}
        T-5: Applied Mathematics and Plasma Physics, Los Alamos National Laboratory, Los Alamos, NM, USA
    }
    \IEEEauthorblockA{\IEEEauthorrefmark{4}
        CAI-2: Computational Physics and Methods, Los Alamos National Laboratory, Los Alamos, NM, USA
    }
     \IEEEauthorblockA{
     email: \{bkkrueger, eidenben, saktar, rbhardwaj, golden, gsgrattan, abhijithj, matsekh, pakin, nsanthi, rtate\}@lanl.gov
    }
}

%

\maketitle

\begin{abstract}

We present the Quantum Hamiltonian Analysis Toolkit (QHAT), a newly developed application that provides a user-friendly interface for studying Hamiltonians and performing Hamiltonian simulation on fault-tolerant quantum computers.  QHAT enables the generation and analysis of Hamiltonians through a powerful and feature-rich application, driven by simple inputs designed to reflect user needs rather than algorithmic details, so that productive research on your application of interest can be done without needing a deep understanding of quantum computing algorithms.

QHAT enables a streamlined workflow to analyze Hamiltonians and Hamiltonian simulation, supporting multiple choices of algorithms and analyses.  It supports Hamiltonians from multiple sources but can also generate Hamiltonians based on a simple description of the system, saving intermediate data files for re-use when generating related Hamiltonians.  Deriving the parameters for quantum computing algorithms can be a challenge, so QHAT is built around user-facing concepts such as maximum allowable error, rather than being built around algorithmic details such as steps counts or order parameters.  An emphasis on user-friendly interfaces and efficient analysis means that the barrier to entry is low while rapidly providing results useful for a broad scope of studies.

\end{abstract}

\begin{IEEEkeywords}
Fault-Tolerant Quantum Computing, Software, Hamiltonian Simulation, Time Evolution, Quantum Phase Estimation, Trotterization, Quantum Singular Value Transform
\end{IEEEkeywords}

\IEEEpeerreviewmaketitle

\section{Introduction}

The Quantum Hamiltonian Analysis Toolkit\footnote{\url{https://github.com/lanl/qhat}} (QHAT or $\hat{Q}$) is a new software package from Los Alamos National Laboratory designed to facilitate studies of quantum computing applications involving Hamiltonians, including determining when or even if such applications would be practical on real-world fault-tolerant quantum computing (FTQC) systems.  QHAT has been developed with the goal of enabling end-to-end analysis of FTQC applications of Hamiltonians, particularly time evolution and phase estimation.  Several important guiding principles have focused the design of QHAT, including:
\begin{enumerate}
    \item{\emph{SME-focused:} The inputs and outputs are chosen to reflect the needs of subject-matter experts (SMEs), such as physicists or material scientists, rather than quantum computing algorithm experts.}
    \item{\emph{User-friendly:} Deep knowledge of quantum computing or of specific application areas should not be required to use QHAT for scientifically interesting work.}
    \item{\emph{Composite design:} The target applications for QHAT have significant re-use of sub-circuits, and any analysis should leverage that re-use to accelerate time to results.}
    \item{\emph{Enabling exploration:} In order to study a wide variety of applications and a wide variety of FTQC approaches, QHAT is written to reflect a conceptual understanding of the options available and to enable easily changing methods or studying the impact of various inputs.}
\end{enumerate}

In order to achieve these goals efficiently, QHAT leverages a variety of available software tools.  QHAT is written in Python and uses open-source packages for quantum chemistry (such as OpenFermion \cite{OpenFermion} and pySCF \cite{pySCF}) and quantum computing (such as Qualtran \cite{Qualtran} and pyLIQTR \cite{pyLIQTR}).

\Cref{sec:related-work} positions QHAT relative to similar software tools.  In \cref{sec:workflow} we discuss an example workflow and how QHAT streamlines the process.  \Cref{sec:qhat_description} expands on that example workflow by discussing more detail about the variety of options available in QHAT\@.  QHAT is actively under development, and \cref{sec:roadmap} discusses the roadmap for development of QHAT, with a significantly expanded set of capabilities.  \Cref{sec:applications} discusses current studies that are enabled by QHAT.

\section{Related Work} \label{sec:related-work}

Other software tools are already available to study quantum computing and/or Hamiltonian analysis.  Relative to QHAT, these tend to fall into two categories: targeted utilities that focus on a subset of the complete workflow and large-scale libraries that provide many capabilities but require expert-level knowledge of each part of the complete workflow.

Targeted utilities include tools that focus on the quantum chemistry at the start of the workflow.  Many such tools exist, including pySCF \cite{pySCF} and OpenFermion \cite{OpenFermion}.  But once a Hamiltonian is defined, the results from these tools then become the input to a completely different set of tools.  In the middle of the workflow, tools such as pyLIQTR provide useful algorithms, but cannot construct a Hamiltonian and have limited analysis capabilities.  Even later in the workflow, tools such as Stim \cite{Stim} focus on performing specific types of algorithmic analysis but provide little support for the construction of complete algorithms that are useful to SMEs.

General-purpose quantum-computing frameworks include Qiskit \cite{Qiskit}, Cirq \cite{Cirq}, PennyLane \cite{PennyLane}, and Azure Quantum/Q\# \cite{Qsharp}.  These can provide many of the desired features but require a detailed understanding not only of programming interfaces but of fine details of each step of the workflow.  For an expert in quantum computing, these are powerful tools.  But for physics, chemistry, or materials SMEs seeking to apply quantum computing to their applications of interest, these are daunting tools that require significant investment to use effectively.

\section{Workflow} \label{sec:workflow}

\begin{figure}[htp]
    \centering
    \includegraphics[width=0.8\columnwidth]{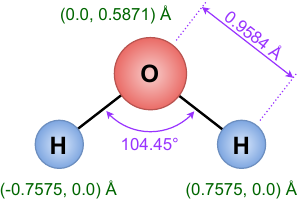}
    \bigskip
    \captionsetup{width=.9\columnwidth}
    \caption{Geometry of our example molecule, water.  Purple data describes the shape of an H\textsubscript{2}O molecule.  Green coordinates are used by QHAT.}
    \label{fig:water}
\end{figure}

QHAT currently provides a simpler, easier-to-use workflow than many other tools rather than providing new capabilities beyond what is available in other tools.  We support this claim by considering an example workflow without QHAT and with QHAT\@.  The workflow we discuss involves the following steps.
\begin{enumerate}
    \item Generate the Hamiltonian in a form that can be simulated on a quantum computer.  We will use a molecular Hamiltonian in this example; see \cref{fig:water}.
    \item Encode the Hamiltonian into a time-evolution operator,
    \begin{equation}
        \op{U}(t) = e^{- i \op{H} t / \hbar} \; .
    \end{equation}
    \item Embed the time-evolution operator into a larger algorithm, in this example a phase estimation algorithm.
    \item Analyze the resulting algorithm.  We will use resource estimation in this example.
\end{enumerate}
Of course many variations of this core workflow exist: other kinds of Hamiltonians, other algorithms, other analyses.  But this example will demonstrate the value of QHAT.

\subsection{Without QHAT}

In order to execute this workflow without QHAT, a user would need to write software around multiple APIs and would need to understand certain ``low-level'' details of how the relevant quantum computing methods work in order to derive appropriate parameter settings.

\subsubsection{Generate the Hamiltonian}  Many tools exist for this process, each of which is designed for different purposes and makes different assumptions.  For a simple molecular Hamiltonian, one choice is to use pySCF and OpenFermion to express the Hamiltonian in second quantization.  This would involve writing software to (1)~call the pySCF API to perform a Hartree-Fock calculation, (2)~write logic to define the active space and call the pySCF API, and (3)~call the OpenFermion API to map the fermionic operators to qubit operators.

\subsubsection{Encode the Hamiltonian} Constructing the time-evolution operator is usually done approximately, for which many methods exist across many different software implementations (such as Qualtran or pyLIQTR)\@.  Using these tools requires writing software against the relevant APIs, but a commonality among most such APIs is the need to set ``low-level'' parameters.  For example, to apply Trotterization for this stage a user would need to select a Trotter formula and derive the correct number of Trotter steps and the correct evolution time.\footnote{Recall that this example is discussing phase estimation, which uses time evolution as a component of a larger algorithm.  While a variety of evolution times could be used, a poor choice can lead to an extremely inefficient final algorithm.}  Selecting these options requires understanding details of Trotterization that are still areas of active research.

\subsubsection{Build the larger algorithm} To embed the time-evolution operator into a larger algorithm requires writing software against one of the various APIs available for this purpose (for example Qualtran or pyLIQTR)\@.  Performing phase estimation requires selecting one of the available methods, then deriving parameter settings.  Knowing how to derive those settings requires a detailed understanding of each method and the approaches for refining the method, such as how the number of phase qubits relates to the evolution time, the energy error, and the probability of failure.

\subsubsection{Analyze the algorithm} Having constructed the full algorithm, some analysis would then be applied to study the algorithm.  For example, resource estimation can be used to approximate the size (in qubits) and runtime (typically approximated by various gate counts) of the algorithm, allowing one to compare algorithms or study what kind of quantum computer would be necessary to run the proposed algorithm.  Once again, this requires writing software to some API.

\subsection{With QHAT}

QHAT simplifies this process by providing a user-focused application instead of software APIs and by deriving appropriate parameter settings based on user-facing options instead of algorithmic details.

\subsubsection{Express the Hamiltonian}  Provide the atoms in the molecule with their coordinates in 3D space.  Specify certain SME-focused details about how to compute the data for this Hamiltonian, such as the desired basis set, the size of the active space, and the fermion-to-qubit operator mapping (the last of which has a ``reasonable'' default).  See \cref{lst:partial_hamgen} for a snippet of a configuration file to specify these options.

\begin{lstlisting}[
    language=Python,
    caption={A partial configuration to build a water molecule Hamiltonian.  See \cref{lst:full_hamgen} for the complete configuration file.},
    label={lst:partial_hamgen},
    float=htbp,
    firstline=13,
    lastline=22,
    firstnumber=13
    ]
hamiltonian.add_atom("O",   0, dy, 0)
hamiltonian.add_atom("H", -dx,  0, 0)
hamiltonian.add_atom("H",  dx,  0, 0)

hamiltonian.basis = "aug-cc-pvtz"

hamiltonian.num_active_occupied = 8
hamiltonian.num_active_vacant = 16

hamiltonian.f2q_mapping = "Jordan-Wigner"
\end{lstlisting}

\subsubsection{Encode the Hamiltonian}  Specify the desired encoding method and certain SME-focused details, such as the maximum allowed error.  Parameter settings are derived and optimized by QHAT, but the user can override any parameter by specifying the desired value.  This allows SME knowledge (such as eigenvalue bounds) or experience with the underlying methods (such as preferring a particular Trotter formula) to take precedence over the default behavior.  See \cref{lst:qre_config_unitary} for a snippet of a configuration file to specify these options.

\begin{lstlisting}[
    language=Python,
    caption={A partial configuration to encode a Hamiltonian as a time-evolution operator.  See \cref{lst:full_qre} for the complete configuration.},
    label={lst:qre_config_unitary},
    float=htbp,
    firstline=12,
    lastline=13,
    firstnumber=12
    ]
unitary.encode_ramped_trotter(
    energy_error = 0.5 * energy_error)
\end{lstlisting}

\subsubsection{Build the larger algorithm}  Specify the desired algorithm and certain SME-focused details, such as the maximum allowed error.  Additional options are derived and optimized to the best of QHAT's ability, but the user can override any derived quantity by specifying the desired value.  See \cref{lst:qre_config_circuit} for a snippet of a configuration file to specify these options.

\begin{lstlisting}[
    language=Python,
    caption={A partial configuration to embed the time-evolution operator in a phase-estimation algorithm.  See \cref{lst:full_qre} for the complete configuration.},
    label={lst:qre_config_circuit},
    float=htbp,
    firstline=16,
    lastline=17,
    firstnumber=16
    ]
circuit.method = "QPE: qualtran textbook"
circuit.energy_error = 0.5 * energy_error
\end{lstlisting}

\subsubsection{Analyze the algorithm}  Specify the desired analysis and (for some analyses) SME-focused parameters.  See \cref{lst:qre_config_analysis} for a snippet of a configuration file to specify these options.

\newpage

\begin{lstlisting}[
    language=Python,
    caption={A partial configuration file to analyze the algorithm.  See \cref{lst:full_qre} for the complete configuration.},
    label={lst:qre_config_analysis},
    float=htbp,
    firstline=20,
    lastline=20,
    firstnumber=20
    ]
analysis.resource_estimator = "pyLIQTR"
\end{lstlisting}

QHAT can execute this entire workflow with two scripts.  First, the user invokes the Hamiltonian generator with the configuration file shown in \cref{lst:full_hamgen}.  This will construct the data file that describes the Hamiltonian.  Second, the user invokes the analysis tool shown in \cref{lst:full_qre}.  This will construct the full algorithm and perform the requested analysis.

\begin{lstlisting}[
    language=Python,
    caption={A complete configuration file to build a Hamiltonian data file for a water molecule.},
    label={lst:full_hamgen},
    float=htbp
    ]
import math

# General configuration
general.print_verbose()
general.file_stub = "water" 
general.file_format = "default" 

# Describe the Hamiltonian
L = 0.9584
theta = math.radians(104.45)
dy = L * math.cos(theta / 2)
dx = L * math.sin(theta / 2)
hamiltonian.add_atom("O",   0, dy, 0)
hamiltonian.add_atom("H", -dx,  0, 0)
hamiltonian.add_atom("H",  dx,  0, 0)

hamiltonian.basis = "aug-cc-pvtz"

hamiltonian.num_active_occupied = 8
hamiltonian.num_active_vacant = 16

hamiltonian.f2q_mapping = "Jordan-Wigner"
\end{lstlisting}

\begin{lstlisting}[
    language=Python,
    caption={A complete configuration file to construct and analyze a complete algorithm.},
    label={lst:full_qre},
    float=htbp
    ]
energy_error = 1.59e-3

# general
general.print_verbose()
general.logfile = "water_TR.qre_log"

# hamiltonian
hamiltonian.load_numpy(
    "water_as-008-016.tensors.npz")

# unitary encoding
unitary.encode_ramped_trotter(
    energy_error = 0.5 * energy_error)

# phase estimation circuit
circuit.method = "QPE: qualtran textbook"
circuit.energy_error = 0.5 * energy_error

# analysis
analysis.resource_estimator = "pyLIQTR"
\end{lstlisting}

\section{QHAT System Description} \label{sec:qhat_description}

While \cref{sec:workflow} outlines a single possible workflow with a single set of options, QHAT has a wider variety of options already available.  An even more extensive set of options is in progress, and will be discussed in \cref{sec:roadmap}.

QHAT separates the processes of Hamiltonian generation from algorithm construction and analysis, because some users prefer to generate their own Hamiltonians using familiar tools.  QHAT can read a Hamiltonian from \texttt{.npz} or \texttt{.hdf5} files, or it can generate a Hamiltonian from a system description.

\subsection{Hamiltonian Generation}

\begin{figure}[ht]
    \centering
    \def\svgwidth{0.75\columnwidth}
    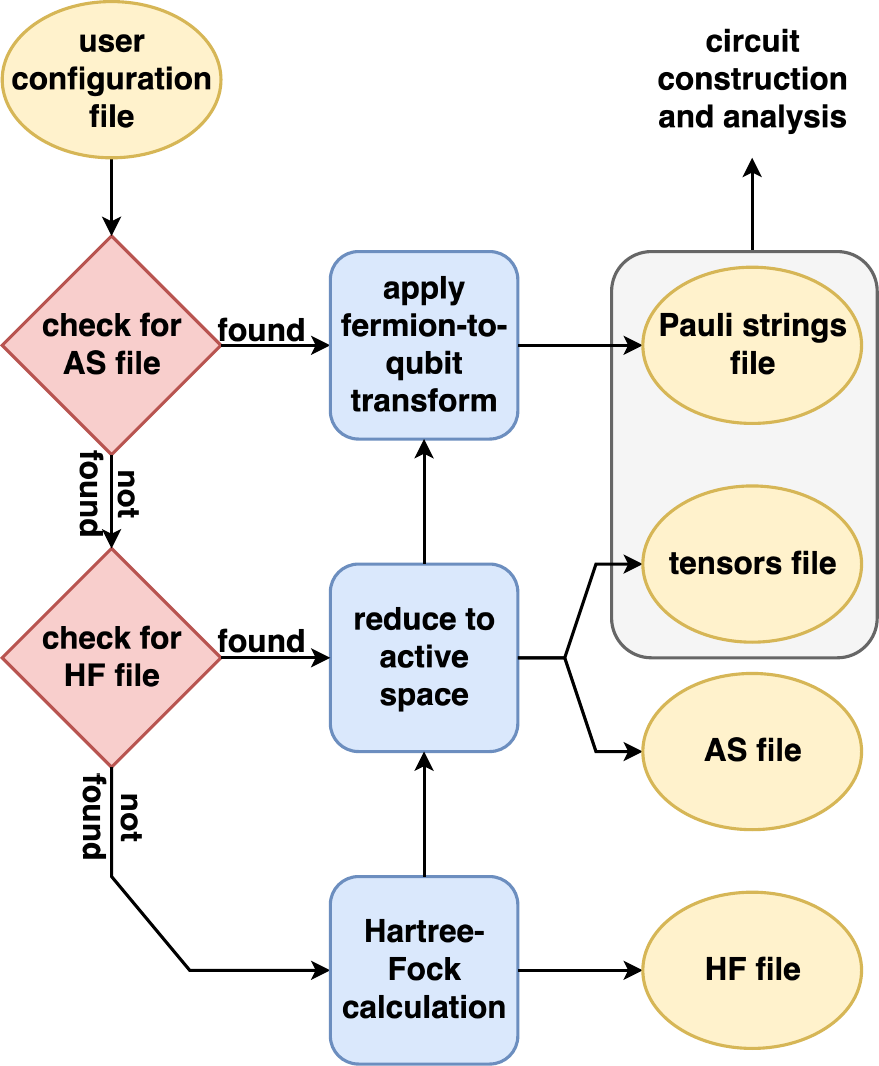
    \captionsetup{width=.9\columnwidth}
    \caption{Diagram of Hamiltonian generator.  Each step further specializes the data, which means that intermediate data can be used to generate related Hamiltonians.}
    \label{fig:hamgen_flowchart}
\end{figure}

QHAT can analyze arbitrary Hamiltonians, but the Hamiltonian generator currently only works for molecular Hamiltonians, and constructs the Hamiltonian in second quantization\footnote{Very briefly: ``first quantization'' focuses on individual particle labels and the quantum states associated with them, while ``second quantization'' focuses on quantum states and tracks how many particles occupy each state.}.  The description of a molecular Hamiltonian consists of
\begin{itemize}
    \item the positions and atomic numbers for each atom in the molecule,
    \item the atomic basis set to use for constructing the molecular basis functions, where basis set data comes from the Basis Set Exchange~\cite{BSE},
    \item the active space of the molecule,\footnote{``Active'' molecular orbitals are able to be used as part of the computation as opposed to being a~priori frozen as occupied or unoccupied.}
    \item the transformation from fermionic operator to qubit operator; Jordan--Wigner and Bravyi--Kitaev are available for use, with Jordan--Wigner being the default.
\end{itemize}
The Hamiltonian generator produces several data files for intermediate states of the calculation so that when it generates a related Hamiltonian the portions of the calculation that are unchanged can be reused instead of recomputed from scratch.  The second-quantization tensors for the molecule are saved in a \texttt{.npz} file so that time-evolution methods that exploit the second-quantization structure can be tested.  The final result, as a set of Pauli strings, is saved as a plain-text file.  Either the second-quantization tensors file or the Pauli strings file can be used as input to the algorithm and analysis part of QHAT\@.  A flowchart of the Hamiltonian generator is shown in \cref{fig:hamgen_flowchart}.

QHAT includes an example configuration file, similar to \cref{lst:full_hamgen}.  These examples demonstrate that the configuration files are Python scripts that can include control flow logic, mathematics expressions, etc., making this a powerful format for specifying complex molecules or performing parameter studies across related molecules.  Also provided is an example script that generates the configuration files for a large number of molecular Hamiltonians (see \cref{sec:hamiltonian_library}).  This script is intended as a demonstration of how to set up many configurations and how to structure the files to enable re-use of shared calculations.  Running this script generates a very large number of configuration files, which would require significant compute resources to then generate and analyze all of the Hamiltonians.

\subsection{Algorithm Construction}

\begin{figure}[ht]
    \centering
    \def\svgwidth{0.95\columnwidth}
    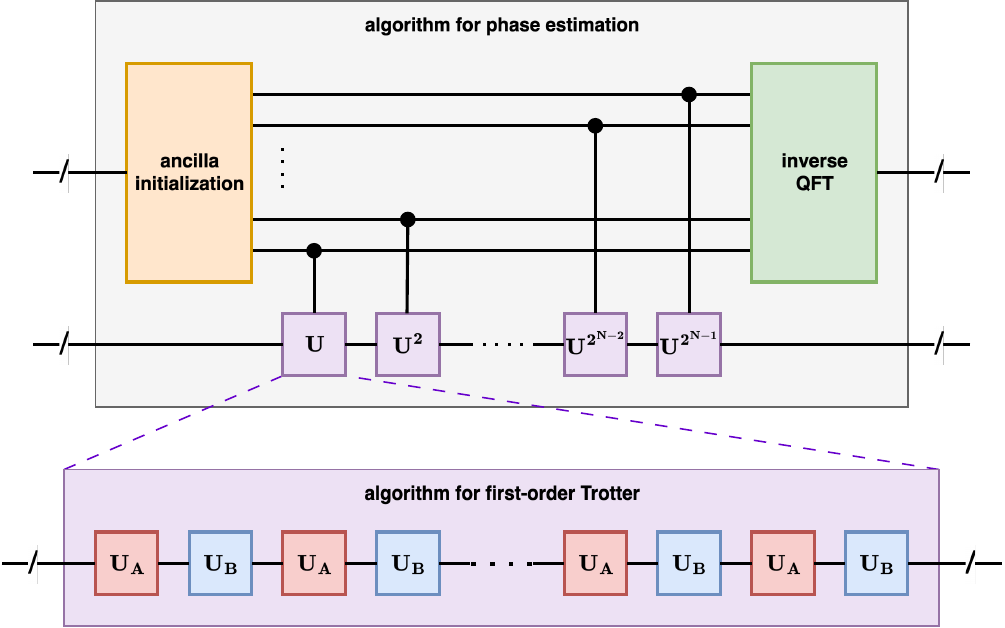
    \captionsetup{width=.9\columnwidth}
    \caption{Diagram of composite algorithm model.  Each red $\op{U}_A$ and blue $\op{U}_B$ is identical so only a single instance of each needs to exist or be analyzed.  The same is true of purple $\op{U}$, except that (a) the $\op{U}$ operator itself is common and a relatively smaller amount of work can extend the analysis to the various different controlled-$\op{U}$ operators, and (b) $\op{U}^p$ is simply $\op{U}$ repeated $p$ times.}
    \label{fig:composite_description}
\end{figure}

QHAT's algorithm-construction capability builds a composite description of a quantum algorithm; see \cref{fig:composite_description} for an example.  The composite nature of this description is important because many of the target applications for QHAT exhibit significant redundancy in their algorithms.  When performing analysis of the algorithms, repeated sub-algorithms can be analyzed a single time and then the results can be duplicated as necessary, which can provide a significant speedup.

QHAT supports multiple Hamiltonian formats for analysis.  Hamiltonians from the QHAT Hamiltonian generator can be loaded into the analysis tool as second-quantization tensors in \texttt{.npz} format or as Pauli strings in plain-text format.\footnote{Certain algorithms leverage the structure provided by the second-quantization tensors so loading the Pauli strings format would preclude using those algorithms.}  Any second-quantization Hamiltonian can be loaded from a \texttt{.npz} file or a \texttt{.hdf5} file.  Any arbitrary Hamiltonian can be loaded as Pauli strings.  These formats are described in the documentation.  Additional readers are added on demand.  Several prototype Hamiltonian readers have been written that were customized to the applications of specific SMEs.  Because they were so tightly coupled to specific applications, they have not been included in QHAT for public use.

QHAT includes time evolution and phase estimation, both of which use the time-evolution operator
\begin{equation} \label{eqn:time_evolution_operator}
    \op{U}(t) = e^{-i \op{H} t / \hbar} \; .
\end{equation}
Methods to construct such an operator typically fall into two categories: product formulas and block encodings.

For product formulas, QHAT provides first- and second-order Trotter expressions.  These use error estimates from Childs et~al.~\cite{Childs_2021} to optimize the choice of Trotter expression and the number of Trotter steps.  This optimization is constrained by a maximum allowable error, which is set by the user.  QHAT's ability to derive and optimize parameters for itself frees the user from having to understand the details of Trotter expressions, error bounds, and other algorithmic details.  Instead, an SME can express the problem in terms that are familiar: approximate the given Hamiltonian to the specified accuracy.

For block encodings, QHAT has a partial, in-progress implementation of double-factorization \cite{DoubleFactorization}, and is studying additional quantum singular value transform (QSVT) methods.  This pathway is not yet as mature as the Trotter pathway, and does not yet do as complete a job of deriving parameters for the user.  However, it is available and provides an alternate approach to constructing the time-evolution operator, which can be experimented with to demonstrate the trade-off space between Trotter and QSVT methods.

QHAT can embed a given time-evolution operator into a larger algorithm.  Currently there are two classes of algorithms: phase estimation and time evolution.

For phase estimation, QHAT provides a ``textbook'' phase estimation algorithm from Qualtran and a ``qubitized'' phase estimation algorithm from pyLIQTR, although other methods are being implemented.  In phase estimation, the ``evolution time'' used in \cref{eqn:time_evolution_operator} needs to be optimized to get a sufficiently accurate answer with the smallest possible quantum circuit.  QHAT performs an initial ``naive'' calculation for this evolution time then optimizes that time to minimize the total circuit size.  It is also necessary to select the number of ancilla qubits for the phase register.  QHAT is able to compute this from the previously-derived ``evolution time'' and user-facing options such as the error constraint and a constraint on the probability of the phase estimation method failing (measuring a value far from the true value, where ``far'' is based on the error constraint).  All derived values can be directly set by the user through simple input options.  Thus, QHAT can enable parameter sweeps and other algorithmic studies while providing sensible values for default behavior.

For time evolution, QHAT provides a ``pure'' time evolution ($\op{U}(t)$ is the full algorithm) and a ``controlled'' time evolution ($\op{U}(t)$ is controlled by a single ancilla qubit).  The former is simple time evolution, to study how a state changes over time; the latter is a building block for phase estimation, and as such is commonly studied to give more insight into larger algorithms.  When studying pure time evolution, users typically will specify an evolution time.  But in both cases, QHAT can provide the optimized evolution time derived for phase estimation.

\subsection{Analysis}

QHAT originated as a resource estimation tool.  Currently, it can analyze the algorithm in terms of Clifford and T gates, reporting the counts of each of these two types of gates as well as the total number of qubits required.  In a typical quantum error-correction code, gates in the Clifford set are very ``inexpensive'' compared to T gates so the total number of T gates is expected to correlate well with the circuit's execution time.  These estimates (particularly qubit and T counts) indicate the capabilities that a fault-tolerant quantum computer would need to have to be able to execute the circuit.  This enables SMEs to gauge what problems they can solve on expected near-term devices, and thereby assess the scientific value that a device could be expected to deliver.  It also enables SMEs to compare different approaches to their problems of interest and experiment with the trade-offs between physical fidelity and computational cost.

QHAT provides numerical simulation as another type of analysis: given an initial state and an algorithm, what is the final state?  While this is practical only for relatively small qubit counts due to the exponential relationship between qubit count and state-vector size, it enables different ways to study Hamiltonian simulation.  If the algorithm in question is phase estimation, numerical simulation can provide eigenvalues and eigenvectors and also probe the probability of failure for the phase estimation method in question.  These enable studies of the error of a given algorithm, the structure of the Hamiltonian, etc.  If the algorithm in question is time evolution, numerical simulation can probe time evolution behavior, error of the translation from a Hamiltonian to a time-evolution operator, and other aspects of the algorithm.

\section{Roadmap for Future QHAT Development} \label{sec:roadmap}

\begin{figure*}[htb]
    \centering
    \def\svgwidth{0.95\textwidth}
    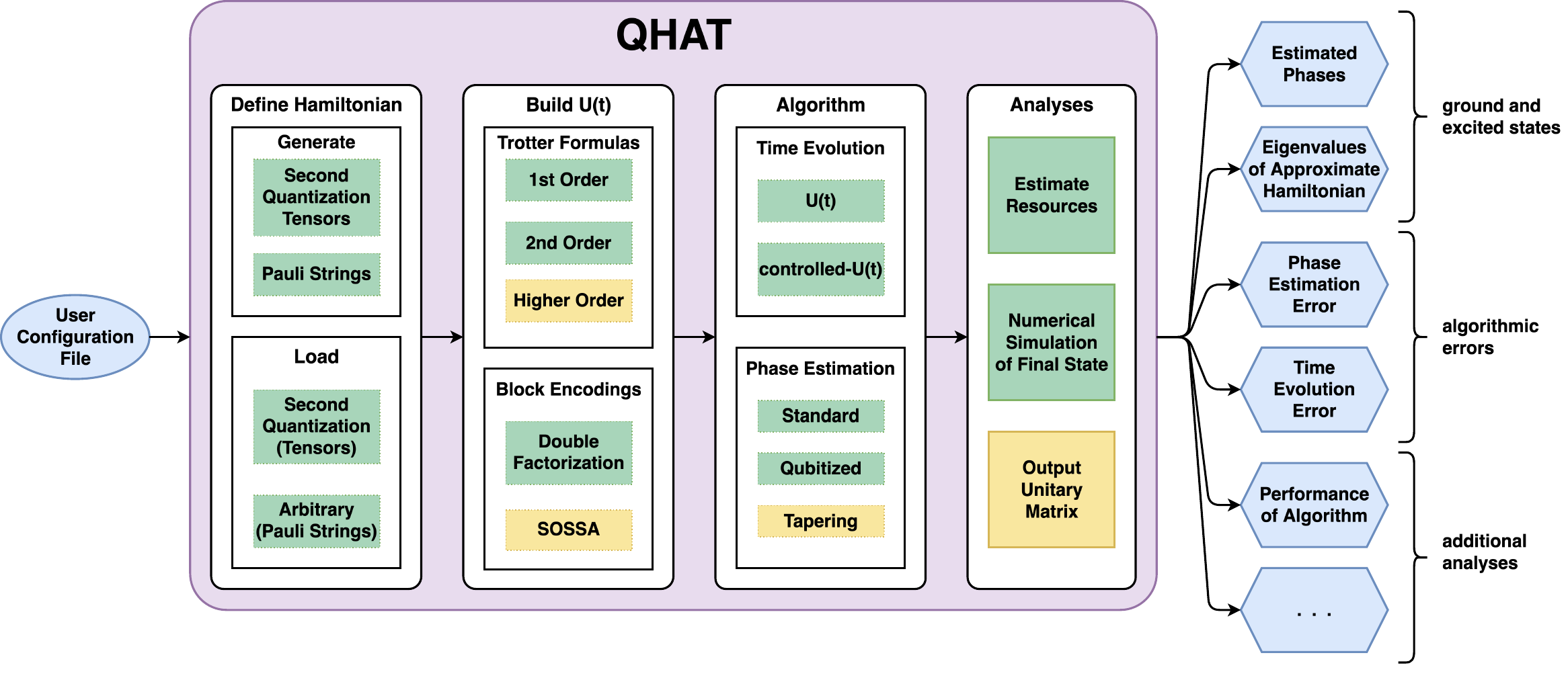
    \caption{Diagram of QHAT workflow, showing the variety of options available (green) and in progress (yellow).  Additional features are in the planning stage, but not shown here.  A simple description of the desired algorithm, with the option to override default behavior, enables QHAT to construct a model and perform the requested analyses.  These results can then be used for further numerical or theoretical study of the Hamiltonian and/or algorithm in question.}
    \label{fig:workflow}
\end{figure*}

\Cref{sec:qhat_description} discussed the current state of QHAT, but active development is in progress for new features, and planning is ongoing for additional features.  Many of these features are expected to be available in the near future.  In addition, user feedback will provide continual guidance on new features that would add significant value.  \Cref{fig:workflow} summarizes the options available, in development, and being planned in QHAT.

\subsection{Hamiltonian Generation} \label{sec:roadmap_hamgen}

The current Hamiltonian generator provides the Hamiltonian, but does not provide any ``initial state'' to use with the algorithms that QHAT can construct.  Work is already in progress on extending the Hamiltonian generator to additionally provide the Hartree-Fock ground state.

Some applications need a more accurate initial state than is provided by the Hartree-Fock method (see, for example, \cref{sec:trotter_error_bounds}).  This suggests that our Hamiltonian generator may need to employ more advanced methods than Hartree-Fock to construct the molecular orbitals and the initial state.

The current Hamiltonian generator works for simple molecules, but many other interesting problems can be studied through Hamiltonian simulation on quantum computers.  We are exploring additional generators, such as Hubbard or Heisenberg Hamiltonians, and are currently working with SMEs on generators for multiple additional types of Hamiltonians, including periodic materials, opacity models, charged-particle stopping power, and nuclear data.

\subsection{Algorithm Construction}

Many algorithms exist for studying Hamiltonians on quantum computers, and the current capabilities of QHAT only just scratch the surface.

QHAT currently only has a small number of file readers to load different kinds of Hamiltonians.
Some applications proposed by SMEs who use QHAT are likely to lead to additional new file readers.  QHAT aims to support a modular design that will enable easily adding new file readers as the need arises.

QHAT currently implements first- and second-order Trotter expressions, but many real-world applications require large Trotter step counts, which provides space for higher-order Trotter expressions to be useful.  Work on fourth-order and higher expressions is in progress.  One of the complications of implementing Trotter expressions into QHAT is the need for bounds on Trotterization error.  QHAT is currently using the results of \cite{Childs_2021} for error bounds, and that analysis needs to be extended to higher-order method.  Additional studies on error bounds such as \cite{Schubert_2023,Burgarth_2024,Hahn_2025} may provide additional ways to estimate the error bounds, and see also \cref{sec:trotter_error_bounds}.

The category of product formulas is larger than just Trotter expressions so QHAT may explore additional product formula approaches, for example qDRIFT \cite{qDRIFT}, in the future.

Product formulas also show a strong dependence on the ordering and grouping of terms in the expressions; see \cref{sec:trotter_ordering}.  QHAT intends to implement schemes for the ordering and grouping of terms in product formulas, enable further optimization of the resulting algorithms.

QHAT's implementation of double-factorization is currently immature and incomplete so there is ongoing work to improve our implementation and expand QHAT's capability to automatically derive reasonable default parameter values for the user.  Many other block encodings exist and there is active work to expand the list of such methods available in QHAT, including methods such as SOSSA \cite{SOSSA1,SOSSA2}.

Once the time-evolution operator $\op{U}(t)$ is constructed, there are various algorithms that can make use of it.  QHAT currently implements, among other algorithms, two phase estimation methods.  Recent studies point to the value of ``high-confidence'' phase estimation methods, such as tapering QPE \cite{taperingQPE}, and there is active work to add one or more of these algorithms to QHAT.

\subsection{Analysis}

Some of the greatest gains in the utility of QHAT are expected to come from adding new types of analysis.

The current resource estimation approaches provide only integrated gate counts for a Clifford+T representation of the circuit.  There could be different gate basis sets beyond Clifford+T that are of interest to users.  Additionally, the ability to analyze gate depth instead of simply integrated gate counts would provide a better estimate of circuit runtime.

Another useful type of analysis is to generate the equivalent unitary matrix for an algorithm and compute its eigendecomposition.  As with numerical simulation, this is only practical for relatively small qubit counts but provides a lot of information about error in the algorithms, ground and excited states (eigenvalues and eigenvectors), and so on.  This type of analysis is being actively developed.

Further types of analysis are possible, depending on the demand from users.  This might include, for example, generating a complete circuit in terms of fundamental gates, or integrations with other tools in the ecosystem of fault-tolerant quantum computing research.

\section{Applications of QHAT} \label{sec:applications}

The development of QHAT is tightly coupled to a number of applications that are actively being studied by ourselves or our collaborators.  The prioritization of features for QHAT is informed by the needs of these studies and other work in progress.  These applications also demonstrate the range of utility for QHAT, including workflows built around answering questions such as
\begin{itemize}
    \item{computing exact ground and excites states (state vectors and/or energies) from eigendecomposition,}
    \item{estimated ground and excited states (state vectors and/or energies) from phase estimation,}
    \item{final states after time evolution,}
    \item{algorithmic error from approximations in building $\op{U}(t)$,}
    \item{error introduced by phase estimation,}
    \item{performance of an algorithm on a quantum computer,}
\end{itemize}
and other potential questions depending on the needs of SMEs.

\subsection{Interactions with Subject-Matter Experts} \label{sec:sme_applications}

We have been collaborating with SMEs on a variety of applications, including materials science, opacity, nuclear data, and charged particle stopping power\footnote{No results are yet published, but a similar effort is documented in \cite{Applications}.}.  This interaction has been useful in understanding how SMEs think about their problems, and in surveying the variety of problems that could be studied using QHAT.  SMEs have been able to estimate resources for their problems of interest to gauge the trade-offs between physical fidelity and the ability to run a circuit on a real-world quantum computer in the foreseeable future.  This has been an important component in identifying applications of fault-tolerant quantum computers that could be practical on early real-world devices and deliver significant scientific value.

\subsection{Hamiltonian Collection} \label{sec:hamiltonian_library}

\begin{figure}[tb]
    \centering
    \includegraphics[width=0.95\columnwidth]{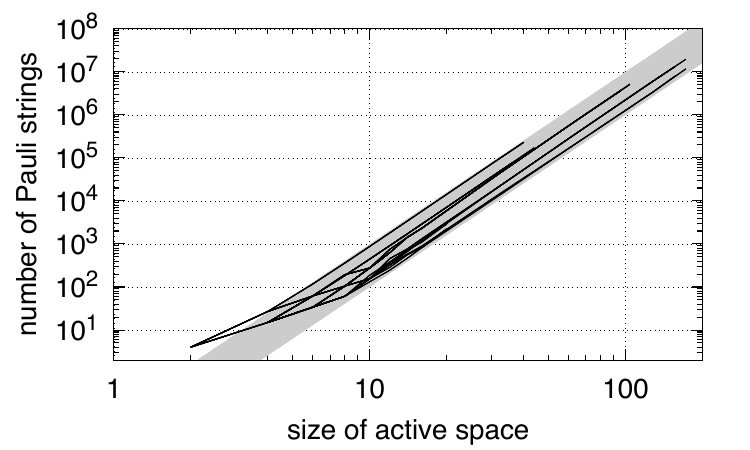}
    \captionsetup{width=.9\columnwidth}
    \caption{Hamiltonians generated by QHAT (\cref{sec:hamiltonian_library}) showing each configuration as a single line where the size of the active space is varied to produce different Hamiltonians.  The shaded region is bounded by $y = 0.01 x^4$ and $y = 0.1 x^4$, showing that the asymptotic increase in the number of Pauli strings is quartic in the size of the active space.  This behavior is consistent with the rank-four tensor that expresses the two-body interactions in a second-quantization Hamiltonian.}
    \label{fig:hamlib_pauli}
\end{figure}

\begin{figure*}[htbp]
  \centering
  \begin{subfigure}[b]{0.95\columnwidth}
    \centering
    \includegraphics[width=0.95\columnwidth]{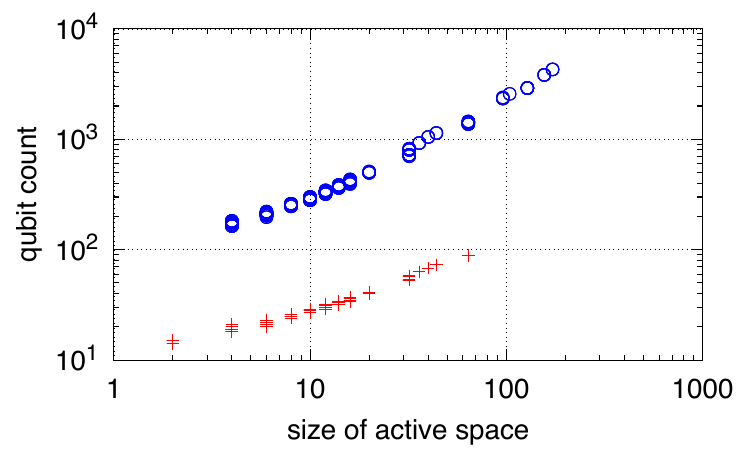}
    \label{fig:top}
  \end{subfigure}
  \begin{subfigure}[b]{0.95\columnwidth}
    \centering
    \includegraphics[width=0.95\columnwidth]{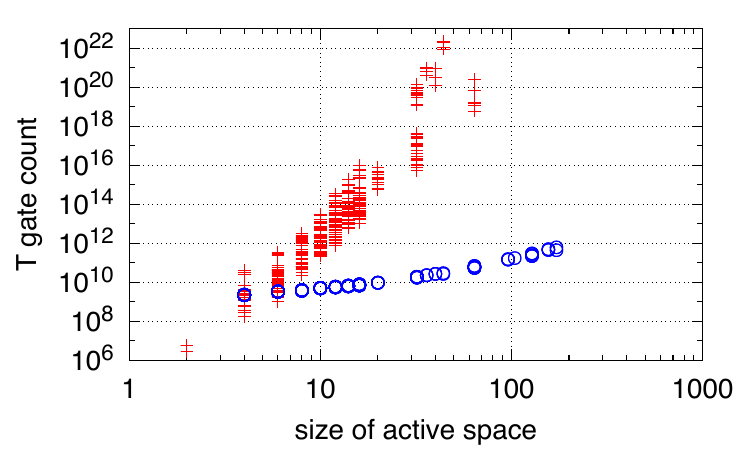}
    \label{fig:bottom}
  \end{subfigure}
  \captionsetup{width=.95\textwidth}
  \caption{Preliminary results for resource estimates of the 540 Hamiltonians shown in \cref{fig:hamlib_pauli}.  The full algorithm is phase estimation; red pluses use Trotterization to approximate \cref{eqn:time_evolution_operator}, while blue circles use double-factorization.  Several important caveats for interpreting this data: (a)~Some models (particularly Trotter with large active spaces and double-factorization with small active spaces) failed to analyze due to bugs that are being fixed.  (b)~The double-factorization implementation is not yet complete so a fixed number of ancilla qubits for phase estimation was chosen rather than letting QHAT set the correct number as with Trotterization; comparing double-factorization to Trotter methods with this data is not a meaningful comparison, but the general trend where double-factorization uses more qubits and less T gates is typical.  (c)~This data uses very little optimization so the gate counts are expected to be much higher than necessary.}
  \label{fig:hamlib_qre}
\end{figure*}

\begin{figure}[htbp]
    \centering
    \includegraphics[width=0.9\columnwidth, trim={0.5cm 1cm 0.5cm 1.75cm}, clip]{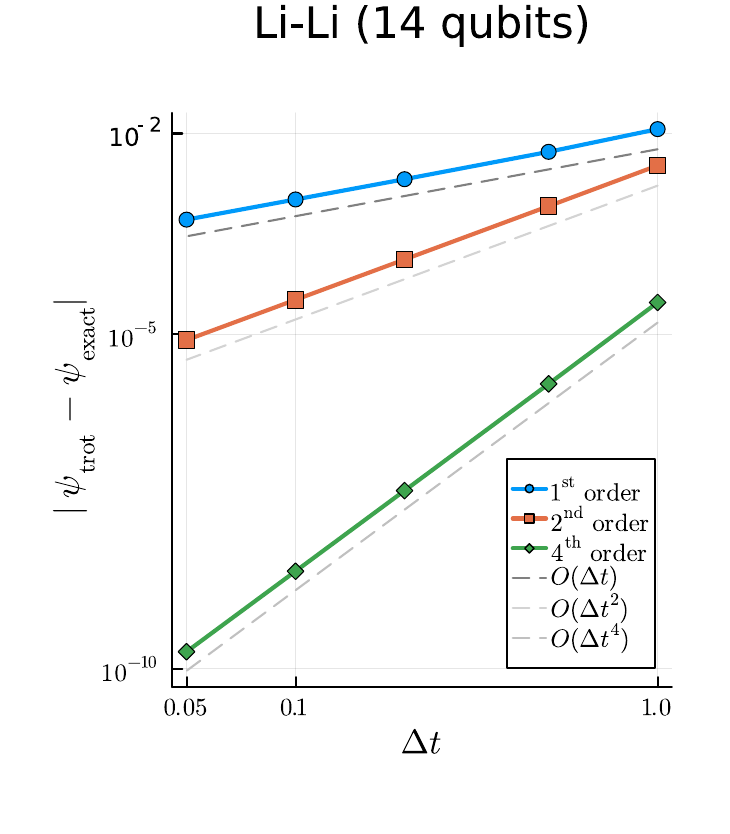}
    \captionsetup{width=.9\columnwidth}
    \caption{Trotterization error in the final time-evolved state for three different Trotter expressions, varying the number of Trotter steps.  This plot uses data for a Li-Li molecule with an active space of 14 molecular orbitals.  The total evolution time is fixed for all calculations.}
    \label{fig:trotter_error}
\end{figure}

\begin{figure}[htb]
    \centering
    \includegraphics[width=0.99\linewidth, trim={0 0 0 1.1cm}, clip]{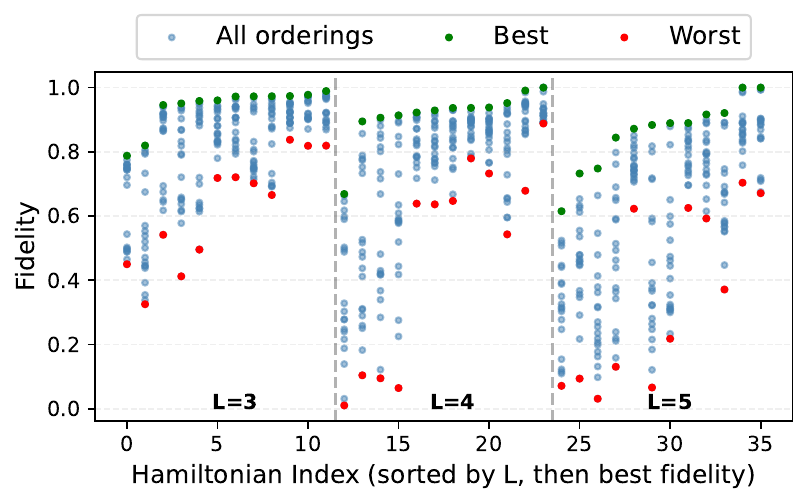}
    \caption{Fidelity of first-order Trotter evolution across randomly sampled term orderings for XXZ Heisenberg Hamiltonians \cite{PhysRevLett.127.037201} with varying system sizes ($L=3,4,5$). Each column corresponds to a single Hamiltonian; blue points show fidelities over all sampled orderings, while green and red points mark the best and worst orderings respectively. The vertical spread illustrates the sensitivity of Trotter error to term ordering.}
    \label{fig:trotter_ordering_fidelity}
\end{figure}

Using the Hamiltonian generation capability, we generated several hundred Hamiltonians using a variant of the example script provided in QHAT for building a suite of configuration files (the range of atomic numbers was much smaller than in the example script).  The collection consists of 540 Hamiltonians, which use 51GB of storage.  It includes:
\begin{itemize}
    \item{two different types of diatomic molecules (homonuclear and hydride);}
    \item{varying atomic number (hydrogen through boron);}
    \item{two different basis sets (STO-6G and HGBS-5);}
    \item{two different interatomic spacings (1\AA\ and the sum of the atomic radii as reported by the \texttt{mendeleev} package);}
    \item{two different fermion-to-qubit operator mappings (Jordan-Wigner and Bravyi-Kitaev);}
    \item{varying active space (remaining as close to 40\% occupied as possible).}
\end{itemize}
These Hamiltonians have provided useful real-world quantum chemistry data for studying various aspects of quantum computing such as those discussed below.  \Cref{fig:hamlib_pauli,fig:hamlib_qre} present some data from this collection.

\subsection{Trotter Expression Error Bounds} \label{sec:trotter_error_bounds}

In order to determine the parameters for a Trotter-based algorithm that satisifes the user-provided error constraints, QHAT needs to compute bounds on the error in Trotter expressions.  Studying the research on these error bounds (e.g., \cite{Childs_2021,Schubert_2023,Burgarth_2024,Hahn_2025}) demonstrated that they remain poorly constrained, with the most detailed studies being performed on Hamiltonians that may behave differently than those of interest to the SMEs we work with.  This motivated a new study of the error of various Trotter expressions, particularly as applied to certain types of Hamiltonians.  This study leverages QHAT to identify preliminary Trotter parameter settings (based on the error bounds from \cite{Childs_2021}).  We hope that the new analyses being added to QHAT will be able to support this study, by generating unitary matrices for numerical analysis and by performing numerical simulation to compute final states, which will help analyze various definitions of the error from Trotterization.  The results of this study will help define new empirical and/or theoretical bounds on Trotter error that can be fed back into QHAT to enable more efficient algorithms.  See \cref{fig:trotter_error} for a sample of partial early results from this study.

Preliminary results from this study suggest that the Hartree-Fock ground state does not always have good overlap with the true ground state, suggesting that we need better initial states for phase estimation.  This points to new features that may need to be developed in QHAT's Hamiltonian generator, in order to design efficient algorithms for Hamiltonians of interest to the SMEs we collaborate with.

\subsection{Term Ordering in Trotter Expressions} \label{sec:trotter_ordering}

Following up on various studies of the effect of changing the ordering and grouping of (commuting) terms in Trotter expressions, QHAT is being used to explore various ordering and grouping strategies \cite{Tate_2026,tranter2019ordering}. This leverages QHAT's ability to implement new logic and test new techniques, along with features discussed above for studying the error in a Trotter expression.  This will enable us to study the effects of term ordering in the context of Hamiltonians of interest to our SMEs.  This study will feed back into QHAT by defining methods to order the terms in QHAT's Trotterization method, enabling QHAT to further optimize algorithms. In addition to supporting currently-existing ordering/grouping strategies in the literature, QHAT will also support custom-defined orderings/groupings. See \cref{fig:trotter_ordering_fidelity}.


\section{Conclusions} \label{sec:conclusions}

QHAT is a new application for analyzing Hamiltonians in the context of fault-tolerant quantum computing.  It is designed to significantly simplify workflows, providing a tool that does not require expert-level knowledge of particular application areas or of quantum computing algorithms in order to productively study quantum computing for Hamiltonian simulation.  A strong baseline of essential features is already in place and providing valuable feedback to our collaborators.  We have also presented a roadmap to a more extensive list of features, many of which are under active development already, that will greatly expand the utility of QHAT.  Several ongoing studies are briefly discussed, indicating the breadth of QHAT's applicability.  QHAT is open-source software available on GitHub\footnote{\url{https://github.com/lanl/qhat}}, and is ready for new users or contributions from interested developers.

\section{Acknowledgments}

The authors would like to thank Carleton Coffrin for productive technical discussions and help setting research directions.  This work was supported by the U.S. Department of Energy through the Los Alamos National Laboratory.  Los Alamos National Laboratory is operated by Triad National Security, LLC, for the National Nuclear Security Administration of U.S. Department of Energy (Contract No. 89233218CNA000001).  The research presented in this article was supported by the NNSA's Advanced Simulation and Computing Beyond Moore's Law Program at Los Alamos National Laboratory. This research used resources provided by the Darwin testbed at Los Alamos National Laboratory (LANL) which is funded by the Computational Systems and Software Environments subprogram of LANL's Advanced Simulation and Computing program (NNSA/DOE).  This work has been assigned LANL technical report number LA-UR-26-23262.

\printbibliography

\end{document}